\documentclass[aps,prl,twocolumn,showpacs]{revtex4}
\usepackage{graphicx}

\begin{document}

\title{Matrix-Product based Projected Wave Functions Ansatz for Quantum Many-Body Ground States}

\author{Chung-Pin Chou$^{1}$, Frank Pollmann$^{2,*}$, and Ting-Kuo Lee$^{1}$}
\affiliation{$^{1}$Institute of Physics, Academia Sinica, Taipei
11529, Taiwan} \affiliation{$^{2}$Max-Planck-Institut f\"{u}r Physik
komplexer Systeme, 01187 Dresden, Germany}

\begin{abstract}
We develop a new projected wave function approach which is based on
projection operators in the form of matrix-product operators (MPOs).
Our approach allows to variationally improve the short range
entanglement of a given trial wave function by optimizing the matrix
elements of the MPOs while the long range entanglement is contained
in the initial guess of the wave function. The optimization is
performed using standard variational Monte Carlo techniques. We
demonstrate the efficiency of our approach by considering a
one-dimension model of interacting spinless fermions. In addition,
we indicate how to generalize this approach to higher dimensions
using projection operators which are based on tensor products.
\end{abstract}


\maketitle


One of the basic problems in physics is to find the ground state of
strongly correlated electron systems. These are of central
importance in several areas of science and technology, including
solid state physics, quantum chemistry as well as nanotechnology.
Analytical exact solutions are only available in few cases and one
thus often relies on numerical studies. Efficient numerical methods
also exist only for special cases. Quantum Monte Carlo (QMC)
simulations \cite{FoulkesRMP01} are for instance only efficient in
the absence of any negative sign problem which occurs in most
fermionic and frustrated systems. By contrast, exact
diagonalizations are applicable in general fermionic models.
However, the size of the corresponding Hilbert space grows
exponentially with the system size and consequently only very small
systems can be considered. In one dimensional systems, very accurate
results have been obtained using the density matrix renormalization
group (DMRG) \cite{SchollwockRMP05}. The DMRG algorithm is closely
related to the concept of matrix-product states (MPS)
\cite{OstlundPRL95,VerstraetePRL04} which can be generalized to
higher dimensions using tensor-product states (TPS)
\cite{SchollwockAP11}. Different TPS based or related approaches
have been introduced recently, e.g., the projected entangled pair
states (PEPS) \cite{Verstraete0407066,VerstraeteAP08}, string-bond
states \cite{SchuchPRL08}, scale-renormalized MPS
\cite{SandvikPRL08}, tensor entanglement renormalization group (TRG)
\cite{ZCGuPRB08,ZCGuPRB09}, correlator-product states
\cite{ChanglaniPRB09,Al-AssamPRB11} and entangled-plaquette states
\cite{MezzacapoNJP09,MezzacapoNJP10,MezzacapoPRB11}. It has been
shown that Monte Carlo sampling in combination with MPS or TPS
yields an efficient method to simulate spin systems
\cite{SandvikPRL07,WangPRB11}. Although there exist applications to
fermionic systems as well, the large amount of entanglement in
fermionic wave functions is very difficult to capture, see e.g.,
Refs.~\cite{KrausPRA10,CorbozPRB10}.

In this Letter, we propose a novel construction of trial wave
functions to efficiently represent ground states of correlated spin
and electronic systems. Our starting point is a known wave function
which we use as a first approximation of the ground state, e.g., a
mean-field wave function or any other better guess we might have. We
then introduce projection operators represented in terms of
matrix-product operators (MPOs)
\cite{verstrate2004,McCulloch2007,Pirvu2010} which are the operator
analogue of MPS. The quality of the trial wave function can then be
successively improved by increasing the dimension of the matrices
used in the MPO. We will refer to these states as
\emph{matrix-product projected states (MPPS)}. Expectation values of
physical observables are computed using Monte Carlo sampling. We use
a variational Monte Carlo (VMC) scheme to optimize the MPO by
minimizing the energy. The proposed MPPS approach can be generalized
to higher dimensions by using tensor-product based
projection-operators. Furthermore, the fermionic structure of the
wave function can be incorporated by using, e.g., a simple Slater
determinant constructed from mean-field Hamiltonians. This approach
can work for frustrated or fermionic systems which cannot be dealt
with using QMC. We discuss the approach for simplicity for the case
of a one-dimensional system in detail and then indicate how to
generalize it to higher dimensions.

\begin{figure}[top]
\rotatebox{0}{\includegraphics[width=3.2in]{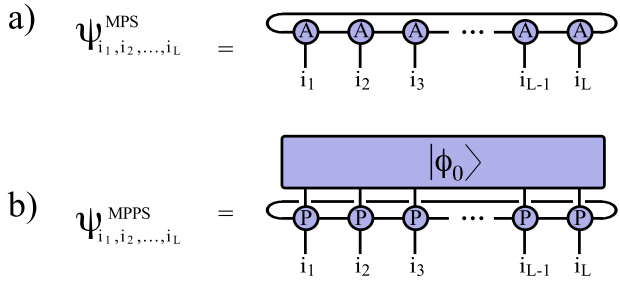}} \caption{(Color
online) Schematic representation of a) a matrix-product state (MPS)
and b) a matrix-product projected state (MPPS) with periodic
boundary condition in a one-dimensional system. $|\phi_{0}\rangle$
is a guess to approximate for the trial wave function
$|\psi\rangle$.}\label{fig1}
\end{figure}

The MPS representation of a quantum state on a chain of length $L$
with periodic boundary condition is given by
\begin{equation}
|\psi^{\text{MPS}}\rangle = \sum_{j_1, \ldots, j_L=1}^{d}
\text{Tr}\left(\hat{A}_{[1]}^{j_1}\hat{A}_{[2]}^{j_2} \dots
\hat{A}_{[L]}^{j_L} \right) |j_1\rangle \ldots |j_L \rangle \label{MPS}
\end{equation}
where $\hat{A}_{[k]}^{j_k}$ are $\chi \times \chi$ matrices. The
physical index $j_k=1\dots d$ represents $d$ local states at site
$k$. This state can be formally rewritten as
\begin{equation}
|\psi^{\text{MPS}}\rangle = \sum_{j_1, \ldots, j_L=1}^{d}
\text{Tr}\left({P}_{[1]}^{j_1}{P}_{[2]}^{j_2} \dots {P}_{[L]}^{j_L}
\right) |\phi_0^{\text{Product}}\rangle \label{MPPS}
\end{equation} in terms of the projector MPOs, ${P}_{[k]}^{j_k}(\equiv\hat{A}_{[k]}^{j_k} |j_k\rangle\langle
j_k|)$, and $|\phi_0^{\text{Product}}\rangle$ is a site-factorized
(product) state defined as
\begin{eqnarray}
|\phi_0^{\text{Product}}\rangle \equiv \prod_{k=1}^L \left(
\sum_{j_k=1}^d |j_k \rangle \right )=\sum_{j_1,\dots,j_L=1}^{d} |
j_1\rangle \dots |j_L \rangle.\label{phi0}
\end{eqnarray}
Clearly, any site-factorized state can be represented using $\chi=1$
matrices which simply renormalize the weight for local states on
each site individually. The entire entanglement in
$|\psi^{\text{MPS}}\rangle$ is expressed by the matrices
$\hat{A}_{[k]}^{j_k}$ and thus a more entangled state requires a
larger $\chi$. The bond dimension $\chi$ needed scales exponentially
with the entanglement entropy $S$, with
$S(=\mbox{Tr}\rho_{\text{red}}\log \rho_{\text{red}})$ being the
von-Neumann entropy of the reduced density matrix
$\rho_{\text{red}}$. The success of the MPS representation in one
dimensional systems is based on the fact that the entanglement in
ground-state wave functions is low and thus they can usually be well
approximated using a small bond dimension $\chi$
\cite{SchollwockAP11}. The main idea of this letter is to choose an
initial wave function $|\phi_0\rangle$ which is already a better
approximation of the ground state than the site-factorized state
$|\phi_0^{\text{Product}}\rangle$ in Eq.~(\ref{MPPS}). The
improvement over an MPS representation is that a part of the
entanglement is contained in $|\phi_0\rangle$. In other words, the
projector MPOs only require modifying the local entanglement since
the long range entanglement is already included in $|\phi_0\rangle$.
This yields a much faster convergence in terms of the bond dimension
compared to MPS (see the example discussed below). Furthermore, it
allows us to express a larger class of states than MPS. In
particular states which violate the area law, e.g., critical states,
can be represented by using the MPPS approach. A pictorial
representation which compares the MPS with the MPPS is given in
Fig.~\ref{fig1}.

As a specific example, we now discuss a model of spinless fermions
at half-filling and show that the trial wave function is
substantially improved over the standard MPS approach by choosing
$|\phi_0\rangle$ to be a mean-field Slater determinant. The
Hamiltonian is given by
\begin{eqnarray}
H=-t\sum_{i}\left(c_{i}^{\dag}c^{\vphantom{{\dag}}}_{i+1}+H.c.\right)+V\sum_{i}\hat{n}_{i}\hat{n}_{i+1},\label{e:eq01}
\end{eqnarray}
where $\hat{n}_{i}=c_{i}^{\dag}c^{\vphantom{{\dag}}}_{i}$ and
$V,t>0$. Using the Jordan-Wigner transformation, the model is
equivalent to the XXZ spin-$1/2$ Hamiltonian on the chain
$H=\sum_{\langle ij\rangle}\left[ \frac{J_{\perp}}{2} \left (s_i^+
s_{i+1}^- + s_i^- s_{i+1}^+\right) + \frac{J_z}{4}
(s_i^z+1)\cdot(s_{i+1}^z+1)\right]$ where $s^{\pm}_i=\frac{s^x_i\pm
i s^y_i}{2}$, $s^{x,y,z}_i$ are the Pauli matrices and are related
to the fermions by $s_i^z=2n_i-1,\,s^+_i=c^\dagger_i,$ and
$s^-_i=c_i$. Thus $J_z=V$ and $J_\perp=-2t$. The discussion below
will be largely in terms of the fermionic model, although all the
results are valid for the spin system as well.

In the non-interacting case (i.e., $V=0$), the Hamiltonian
(\ref{e:eq01}) has a simple ground-state wave function which we
choose as the initial wave function for the MPPS:
\begin{eqnarray}
|\phi_{0}^{\text{Slater}}\rangle=\prod_{\vec{k}}^{\epsilon_{\vec{k}}<\epsilon_{F}}c_{\vec{k}}^{\dag}|0\rangle=\sum_{j_1,
\ldots, j_L}S_{j_1, \ldots, j_L}|j_1\rangle \dots |j_L\rangle.
\label{e:eq06}
\end{eqnarray}
Here $\epsilon_{F}$ denotes the Fermi energy, $\epsilon_{\vec{k}}$
is the dispersion relation and the local states are given by
$j_k=0,1$. $S_{j_1, \ldots, j_L}$ is the Slater determinant for the
configuration $|j_1\rangle \ldots |j_L\rangle$. Using this initial
wave function, the MPPS Ansatz reads
\begin{eqnarray}
|\psi^{\text{MPPS}}\rangle&=&\sum_{j_1, \ldots, j_L}\text{Tr}\left({P}_{[1]}^{j_1}\dots{P}_{[L]}^{j_L} \right) |\phi_0^{\text{Slater}}\rangle\label{e:eq07}\\
&=&\sum_{j_1, \ldots,
j_L}\text{Tr}\left(\hat{A}_{[1]}^{j_{1}}\cdots\hat{A}_{[L]}^{j_{L}}\right)S_{j_1,
\ldots, j_L}|j_1\rangle \dots  |j_L\rangle.\nonumber
\end{eqnarray}
There are two points in parameter space at which the ground-state
wave function can be represented exactly with $\chi=1$. Firstly, if
$V=0$, the ground state is given by $|\phi_0^{\text{Slater}}\rangle$
and thus does not require any modifications (i.e.,
$\hat{A}^{j_k}_{[k]}=1$ for all $k$ and $j_k$). Secondly, if
$V\rightarrow\infty$, the ground state is a two-fold degenerate
charge density wave (CDW) state (which is a product state in this
limit). This state is simply obtained by choosing for example
$\hat{A}^{1}_{[2k]}=\hat{A}^{0}_{[2k+1]}=1$ for all $k$ (all other
elements being zero) -- everything except one of the two CDW states
is projected out from the Slater determinant. Note that in the
former case, the state has algebraic correlations and thus an MPS
requires a bond dimension which roughly scales linearly with the
system size \cite{PollmannPRL09}. In the latter case, the MPS can
also represent the ground state with $\chi=1$.

Additionally, we consider for comparison another popular variational
state with density-density Jastrow correlation factor. The
Jastrow-type wave function can be defined as
\begin{eqnarray}
|\psi^{\text{Jastrow}}\rangle=\exp\left(\sum_{i<j}\eta_{ij}\hat{n}_{i}^{h}\hat{n}_{j}^{h}\right)|\phi_{0}^{\text{Slater}}\rangle\label{e:eq09}
\end{eqnarray}
with
$\eta_{ij}\equiv\ln\left(r_{ij}^{\alpha}v_{\beta}^{\delta_{j,i+\beta}}\right)$.
Here
$r_{ij}=\left|\sin\left(\frac{\pi}{L}(x_{i}-x_{j})\right)\right|$
and $\hat{n}_{i}^{h}=1-\hat{n}_{i}$. The two parameters $v_{\beta}$
with the nearest and second-nearest neighbor indices $\beta$ are for
short range hole-hole repulsion if these values are less than $1$.
The factor $r_{ij}^{\alpha}$ is for long range correlations and
repulsive if $\alpha$ is positive. In fact, we could use the
optimized $|\psi^{\text{Jastrow}}\rangle$ as the initial wave
function for the MPPS approach (instead of
$|\phi_{0}^{\text{Slater}}\rangle$). However, here we use
$|\psi^{\text{Jastrow}}\rangle$ only as a benchmark to compare to
the Slater determinant based MPPS.

We now use the VMC method to optimize the energy using the three
different trial wave functions (MPS, MPPS and Jastrow) and to
compute their correlation functions as well. The energy optimization
is based on the stochastic reconfiguration (SR) method developed in
order to optimize many parameters \cite{SorellaPRB01}. To stabilize
the SR method, we also use a truncation technique for irrelevant
variational parameters \cite{CasulaJCP04} and a stochastic annealing
approach \cite{SandvikPRL07}. The scaling of the numerical
complexity resulting from calculating the matrix product scales as
$\chi^3$. Note that all the matrix elements in
$\hat{A}_{[k]}^{j_{k}}$ of the MPPS could be the variational
parameters to be optimized. However, for simplicity, we assume the
matrices to be real, symmetric and translationally invariant here.

\begin{figure}[top]
\rotatebox{0}{\includegraphics[height=4.2in,width=3.5in]{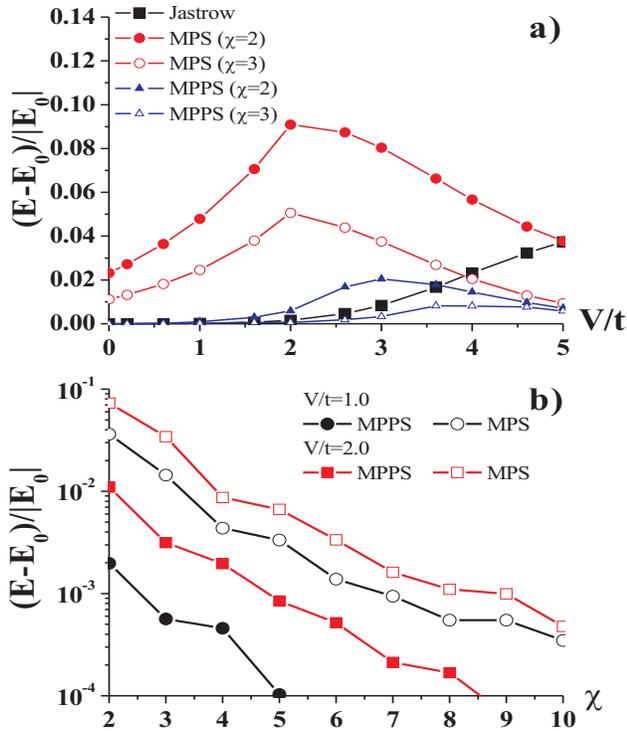}}
\caption{(Color online) a) Comparison of the optimized energies for
different trial states with exact diagonalization results for the
$t-V$ model on a $L=18$ chain with periodic boundary condition. b)
Comparison of the optimized energies with infinite time-evolving
block decimation results for different bond dimensions $\chi$ using
the MPS and the MPPS for $V/t=1,2$ on a $L=202$ chain. $E_{0}$ is
the exact ground-state energy.}\label{fig2}
\end{figure}


Figure~\ref{fig2}a compares the optimized energies for Hamiltonian
(\ref{e:eq01}) using a $L=18$ chain with periodic boundary condition
with the exact ground-state energy $E_{0}$. The exact ground-state
energies are obtained using sparse matrix diagonalization. The
advantage of the MPPS approach over the regular MPS is especially
clear in the regime of small $V/t$. Here, the $V=0$ ground state
used in the MPPS is a much better approximation of the ground state
compared to the site-factorized wave function used in the MPS. Thus
it pays off that part of the entanglement is already contained in
$|\phi_0^{\text{Slater}}\rangle$ making the MPPS more efficient. The
MPPS yields lower energies in the regime of large $V/t$ as well,
even though the advantage is less pronounced. The simple
density-density repulsive Jastrow-type wave function is a good
approximation of the ground states for small $V/t$, however, it
fails to describe the gapped phases in the regime $V/t>2$. In
particular, it does not capture the phase transition to the charge
ordered state at $V/t=2$. Figure~\ref{fig2}b shows the
$\chi$-dependence of the energies and compares the results to the
exact ground-state energies $E_{0}$ in the thermodynamic limit. The
energies for infinite systems are obtained using a infinite
time-evolving block decimation (iTEBD) \cite{VidalPRL07} algorithm
using MPS with large bond dimensions ($\chi>300$). In the gapless
phase ($V/t=1$), we find that the MPPS results show much faster
convergence to the ground state than the MPS. The MPPS with $\chi=5$
already has an energy which is very close to the ground state (error
off $10^{-4}$) while the MPS displays comparably slow convergence up
to $\chi=10$. Even at the critical point ($V/t=2$), the MPPS gives
significantly better results than the MPS.

\begin{figure}[top]
\rotatebox{0}{\includegraphics[height=4.2in,width=3.5in]{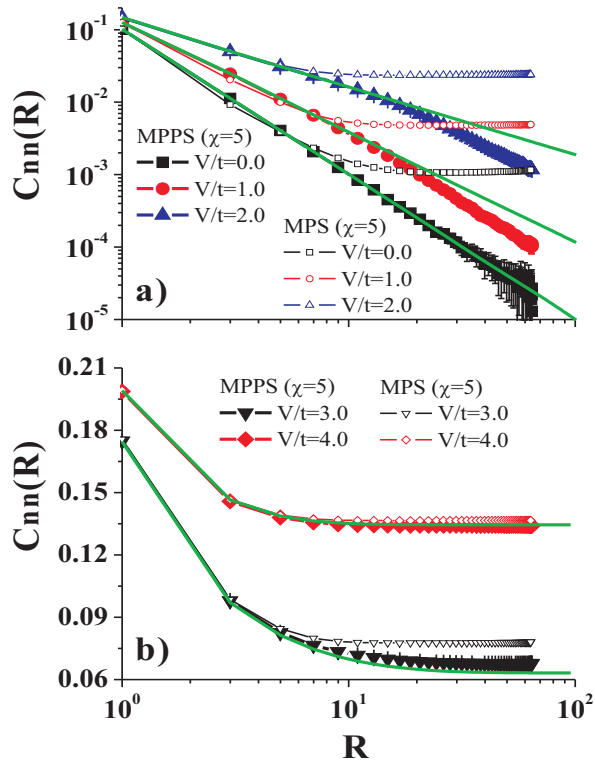}}
\caption{(Color online) Density-density correlation function
$C_{nn}(R)$ for the MPPS and the MPS with the same bond dimension as
a function of distance $R$ in a) gapless and b) CDW regions on a
$L=202$ chain. The green lines are obtained from the MPS with a
large bond dimensions ($\chi>300$) on infinite chains. For the chain
with a finite length, the periodic boundary condition has been
considered in the distance $R$.}\label{fig3}
\end{figure}

In order to further test the quality of the wave function, we
compute the density-density correlation function
\begin{eqnarray}
C_{nn}(R)=\frac{1}{L}\sum_{i}\langle n_{i}n_{i+R}\rangle-\frac{1}{4}.\label{e:eq08}
\end{eqnarray}
Figure~\ref{fig3} shows the obtained correlation function $C_{nn}$
for the MPS and the MPPS using the same bond dimension with the
exact results. The exact results are calculated using the iTEBD
algorithms with large bond dimensions ($\chi>300$). Note that to
avoid the staggered decay of the correlation function arising from
the $2k_{F}$ components, we only show the correlation function with
odd sites. Let us first discuss the gapless regime shown in
Fig.~\ref{fig3}a. In the $V=0$ case, the MPPS results are exact (up
to the statistical errors) and we obtain the expected $R^{-2}$ decay
of the correlations. For finite $V/t$, the exponents change and the
correlation function is analytically given by
$C_{nn}(R)=\frac{A}{R^{2}}+(-1)^{R}\frac{B}{R^{2\kappa}}$ where $A$
and $B$ are non-universal constants and
$\kappa^{-1}=\frac{2}{\pi}\arccos\left(-V/2t\right)$
\cite{GiamarchiBOOK04}. The MPPS results agree with the exact
algebraic decay over $\sim$$50$ lattice sites while the regular MPS
captures only about $\sim$$5$ lattice sites (using the same bond
dimension). In order to understand the fast convergence for the MPPS
in Figure~\ref{fig2}b, it is interesting to compare the behavior of
the MPS and the MPPS at long distances in Figure~\ref{fig3}a. The
MPS contains no long range entanglement and thus the correlation
functions saturates to a constant as $R$ goes to infinity. In the
MPPS, we start from a free fermion wave function in which the local
entanglement is modified by the matrix-product projection-operators.
Thus the correlation functions fall of $\propto R^{-2}$ at long
distances. This shows us intuitively why the MPPS converges faster
in terms of the bond dimension $\chi$: The matrices do not need to
transform a product state into one with algebraic correlation but
they only need to modify the exponent of a wave function which is
already critical. In Fig.~\ref{fig3}b, we also present the
correlation function in the gapped CDW region. The CDW order
parameters, enhanced as increasing $V/t$, can be estimated from the
converged values of the correlation functions at large distance. In
addition to the resemblance on the correlation functions between the
MPPS and exact results, the MPPS also obtains a better estimation
for the order parameter than the MPS with the same bond dimension.

The natural extension of the MPPS to two dimensions is to use
tensor-product projected states (TPPS). The trial wave function is
then formally written identical to Eq.~(\ref{MPPS}) with the
amplitudes $W_{j_1,j_2,\ldots,j_N}$ of the projections given by a
tensorial trace \cite{VerstraetePRL06,ShiPRA06}:
\begin{eqnarray}
W_{j_1,j_2,\ldots,j_N} =
\text{tTr}\left(\hat{T}_{[1]}^{j_{1}}\hat{T}_{[2]}^{j_{2}}\cdots
\hat{T}_{[N]}^{j_{N}}\right).\label{e:eq10}
\end{eqnarray}
Here $\hat{T}_{[k]}^{j_{k}}$ are rank $z$-tensors with $z$ being the
coordination number of the lattice and $N$ lattice size. The
tensorial trace tTr is a generalization of the matrix-trace and
implies a contraction over all virtual bond indices. The main
difficulty is that the tensorial trace cannot be exactly evaluated
anymore. This is because the numerical effort to contract the
tensors grows exponentially with the linear dimension of the
systems. To overcome this problem, we can utilize several
approximate ways to calculate the contraction
\cite{JiangPRL08,GuPRB08}. One way to approximate the trace is to
use the tensor entanglement renormalization group method
\cite{GuPRB08}. The numerical complexity depends then on the
lattice, e.g., applied to a honeycomb lattice, and that of the
approximate evaluation of the tensorial trace scales as $\chi^6$.
The generalization of the method to two dimensions is now being
studied \cite{eddy2012}.


In conclusion, we have introduced a projected wave function approach
which is based on matrix-product projection-operators. The approach
is as a generalization of the MPS in which the initial state is not
necessarily assumed to be close to a site-factorized state. We have
demonstrated that the quality of the wave functions, MPPS, is
superior to the MPS with the same bond dimension by comparing
energies and correlation functions for a one-dimension model of
spinless fermions. The approach can be generalized to higher
dimensions by using tensor-product based projection-operators. Here
we expect our approach to be particularly useful because the
computational complexity grows very quickly with the dimension of
the tensors and thus it is very beneficial to express part of the
entanglement by using a suitable initial wave function.

\begin{acknowledgments}
The authors would particularly like to acknowledge discussions with
A. Turner, K. Penc, and P. Fulde. CPC acknowledge funding from NSC
in Taiwan with Grant No. 98-2112-M-001-017-MY3 and MPIPKS in
Germany. The simulations were performed in the National Center for
High-performance Computing and the PC Cluster III of Academia Sinica
Computing Center in Taiwan. The calculations were partially
performed on the PC Clusters of MPIPKS in Germany.
\end{acknowledgments}

$^{*}$ frankp@pks.mpg.de


\begin{thebibliography}{100}
\bibitem{FoulkesRMP01} W. M. C. Foulkes \textit{et al.}, Rev. Mod. Phys. \textbf{73}, 33 (2001).
\bibitem{SchollwockRMP05} U. Schollw\"{o}ck, Rev. Mod. Phys. \textbf{77}, 259 (2005).
\bibitem{OstlundPRL95} S. Ostlund and S. Rommer, Phys. Rev. Lett. \textbf{75}, 3537 (1995).
\bibitem{VerstraetePRL04} F. Verstraete, D. Porras, and J. I. Cirac, Phys. Rev. Lett. \textbf{93}, 227205 (2004).
\bibitem{SchollwockAP11} U. Schollwock, Ann. of Phys. \textbf{326}, 96 (2011).
\bibitem{Verstraete0407066} F. Verstraete and J. I. Cirac, arXiv:cond-mat/0407066.
\bibitem{VerstraeteAP08} F. Verstraete, V. Murg, and J. I. Cirac, Adv. Phys. \textbf{57}, 143 (2008).
\bibitem{SchuchPRL08} N. Schuch, M. M. Wolf, F. Verstraete, and J. I. Cirac, Phys. Rev. Lett. \textbf{100}, 040501 (2008).
\bibitem{SandvikPRL08} A. W. Sandvik, Phys. Rev. Lett. \textbf{101}, 140603 (2008).
\bibitem{ZCGuPRB08} Z.-C. Gu, M. Levin, and X.-G. Wen, Phys. Rev. B \textbf{78}, 205116 (2008).
\bibitem{ZCGuPRB09} Z.-C. Gu, M. Levin, B. Swingle, and X. G. Wen, Phys. Rev. B \textbf{79}, 085118 (2009).
\bibitem{ChanglaniPRB09} H. J. Changlani, J. M. Kinder, C. J. Umrigar, and GarnetKin-Lic Chan, Phys. Rev. B \textbf{80}, 245116 (2009).
\bibitem{Al-AssamPRB11} S. Al-Assam, S. R. Clark, C. J. Foot, and D. Jaksch, Phys. Rev. B \textbf{84}, 205108 (2011).
\bibitem{MezzacapoNJP09} F. Mezzacapo \textit{et al.}, New. J. Phys. \textbf{11}, 083026 (2009).
\bibitem{MezzacapoNJP10} F. Mezzacapo and J. I. Cirac, New. J. Phys. \textbf{12}, 103039 (2010).
\bibitem{MezzacapoPRB11} F. Mezzacapo, Phys. Rev. B \textbf{83}, 115111 (2011).
\bibitem{SandvikPRL07} A. W. Sandvik and G. Vidal, Phys. Rev. Lett. \textbf{99}, 220602 (2007).
\bibitem{WangPRB11} L. Wang, I. Pi\v{z}orn, and F. Verstraete, Phys. Rev. B \textbf{83}, 134421 (2011).
\bibitem{KrausPRA10} C. V. Kraus, N. Schuch, F. Verstraete, and J. I. Cirac, Phys. Rev. A \textbf{81}, 052338 (2010).
\bibitem{CorbozPRB10} P. Corboz, R. Orus, B. Bauer, and G. Vidal, Phys. Rev. B \textbf{81}, 165104 (2010).
\bibitem{verstrate2004} F. Verstraete, J.J. Garcia-Ripoll and J.I. Cirac, Phys. Rev. Lett. \textbf{93} 207204 (2004).
\bibitem{McCulloch2007} I.P. McCulloch, J. Stat. Mech.: Theor. Exp. P10014 (2007).
\bibitem{Pirvu2010} B. Pirvu, V. Murg, J.I. Cirac and F. Verstraete, New J. Phys. \textbf{12}, 025012 (2010).
\bibitem{PollmannPRL09} F. Pollmann, S. Mukerjee, A. M. Turner, and J. E. Moore, Phys. Rev. Lett. \textbf{102}, 255701 (2009).
\bibitem{SorellaPRB01} S. Sorella, Phys. Rev. B \textbf{64}, 024512 (2001).
\bibitem{CasulaJCP04} M. Casula, C. Attaccalite, and S. Sorella, J. Chem. Phys. \textbf{121}, 7110 (2004).
\bibitem{VidalPRL07} G. Vidal, Phys. Rev. Lett. \textbf{98}, 070201 (2007).
\bibitem{GiamarchiBOOK04} T. Giamarchi, "\textit{Quantum Physics in One Dimension}", (Oxford, 2004).
\bibitem{VerstraetePRL06} F. Verstraete, M. M. Wolf, D. Perez-Garcia, and J. I. Cirac, Phys. Rev. Lett. \textbf{96}, 220601 (2006).
\bibitem{ShiPRA06} Y.-Y. Shi, L.-M. Duan, and G. Vidal, Phys. Rev. A \textbf{74}, 022320 (2006).
\bibitem{JiangPRL08} H. C. Jiang, Z. Y. Weng, and T. Xiang, Phys. Rev. Lett. \textbf{101}, 090603 (2008); Z. Y. Xie, H. C. Jiang, Q. N. Chen, Z. Y. Weng, and T. Xiang, \textit{ibid.} \textbf{103}, 160601 (2009).
\bibitem{GuPRB08} Z.-C. Gu, M. Levin, and X.-G. Wen, Phys. Rev. B \textbf{78}, 205116 (2008); Z.-C. Gu, M. Levin, B. Swingle, and X. G. Wen, \textit{ibid.} \textbf{79}, 085118 (2009).
\bibitem{eddy2012} C.-P. Chou, F. Pollmann, and T.-K. Lee, in preparation (2012).
\end{thebibliography}
\end{document}